\documentclass[pdftex,twocolumn,epjc3]{svjour3}          % twocolumn

\RequirePackage{fix-cm}
\RequirePackage[T1]{fontenc}

\smartqed  % flush right qed marks, e.g. at end of proof

\RequirePackage{graphicx}
\RequirePackage{mathptmx}      % use Times fonts if available on your TeX system
\RequirePackage{flushend}
\RequirePackage[numbers,sort&compress]{natbib}
\RequirePackage[colorlinks,citecolor=blue,urlcolor=blue,linkcolor=blue]{hyperref}

\usepackage{eurosym}
\usepackage{amssymb}
\usepackage{amsfonts}
\usepackage{amsmath}
\usepackage{mathtools}

\journalname{Eur. Phys. J. C}

\begin{document}

\title{Exact modifications on a vacuum spacetime due to a gradient bumblebee field at its vacuum expectation value}

\author{F. P. Poulis\thanksref{addr1,e1}
        \and
        M. A. C. Soares\thanksref{addr2,e2} %etc.
}

\thankstext{e1}{e-mail: fp.poulis@ufma.br (corresponding author)}
\thankstext{e2}{e-mail: soares.mauricio@discente.ufma.br}

\institute{Coordena\c{c}\~{a}o do Curso Interdisciplinar em Ci\^{e}ncia e Tecnologia, Universidade Federal do Ma\-ra\-nh\~{a}o, S\~{a}o Lu\'{\i}s, Maranh\~{a}o, 65080-805, Brazil\label{addr1}
\and
Departamento de F\'{\i}sica, Universidade Federal do Ma\-ra\-nh\~{a}o, S\~{a}o Lu\'{\i}s, Maranh\~{a}o, 65080-805, Brazil\label{addr2}
}

\date{Received: 3 January 2022 / Accepted: 23 June 2022}

\maketitle

\begin{abstract}
This work belongs to the context of the standard-model extension, in which a Lorentz symmetry violation is induced by a bumblebee field as it acquires a nonzero vacuum expectation value. The mathematical formulation of a generic bumblebee model and its associated dynamical equations are presented. Then, these equations are considered for the vacuum and a substantial simplification is performed for the particular case of a gradient bumblebee field at its vacuum expectation value. After some further manipulation, a method to easily find solutions to the model is developed, in which the exact effect on the spacetime description due to the presence of this bumblebee field is explicitly provided. As some examples, the method is applied to determine the implications of the bumblebee field on the Schwarzschild spacetime and also on a rotating one. A previously published solution is recovered and some new ones are obtained. In the rotating situation, a simple solution is found which contains both the Kerr solution and the already published one as special cases. It is also shown its distinguished surfaces are still given by the same corresponding expressions for the Kerr solution. In conclusion, the mathematical improvement made is considered to be a significant contribution to the theory as a powerful tool to investigate its many aspects and consequences.
\end{abstract}

\section{Introduction}
\label{Intro}

Gravity is the only fundamental interaction devoid of a successfully verified quantum formulation. Even though there are many proposals \cite{Carlip:2015asa}, none of them can be directly tested, since the typical effects of quantum gravity take place at the Planck scale, whose energy level ($\sim 10^{19}$ GeV) is far beyond the reach of current experiments. Nevertheless, there can also be some indirect consequences of these effects that are more likely to be observed and would certainly shed some light on this pursuit if that happened.

Several quantum gravity models lead to the violation of Lorentz symmetry in the Planck regime. As examples, we can cite the string field theory \cite{Kostelecky1988zi,Kostelecky1989jp,Kostelecky1989jw,Kostelecky1991ak}, noncommutative field theories \cite{Carroll2001ws, Mocioiu2000ip,Ferrari2006gs}, and loop quantum gravity \cite{Gambini1998it,Ellis1999uh}, among others \cite{Berger2001rm,Blas2014ira01,Blas2014ira02,Lehnert2013axa,Gorbunov2005dd}. This violation, in turn, also contributes to the spacetime dynamics on lower energy levels and, thus, can produce astronomical consequences. In other words, the breaking of Lorentz symmetry in the Planck scale may ultimately have implications on every solution from General Relativity (GR), thereby leaving its traces in possibly any astronomical phenomenon.

Therefore, it is of great interest to verify whether or not these implications agree with the observational data as they could indicate the existence of a certain quantum gravity model. To do so, however, one must have the appropriate description of the observed phenomenon in the context of the model under investigation and this, naturally, requires the corresponding spacetime solution that takes into account the low energy contributions associated with the Lorentz symmetry violation. It is, then, essential to determine how these contributions will affect spacetime, in special, what modifications it will have as compared to its description according to GR.

As a general theoretical framework for testing Lorentz symmetries, there is an effective field theory, named {\em stan\-dard-model extension\/} (SME) \cite{Kostelecky1994rn,Colladay1996iz,Colladay1998fq,Kostelecky2003fs}, that describes the\linebreak standard model of particle physics coupled with GR whose mathematical structure includes terms that violate Lorentz symmetry at the Planck scale. It constitutes an intense research line for decades, with many studies already made on its electromagnetic \cite{Colladay2001wk,Kostelecky2000mm,Bakke2014kua,Kostelecky1999zh,Yoder2012ks,PhysRevD.68.085003,Lehnert2004ri,Kharlanov2007yp,Kostelecky2001mb,Kostelecky2002hh,Kostelecky2006ta,Carroll1989vb,Adam2001ma,Adam2002rg,Moyotl2013tva,Chen2000zh,Carone2006tx,Klinkhamer2010zs,Schreck2011ai,Hohensee2008xz,Altschul2005mu}, electroweak \cite{Colladay2009rb,Mouchrek-Santos2016upa}, and strong \cite{PhysRevLett.108.261603} sector, besides the hadronic physics \cite{Berger2015yha} as well. Moreover, some of its gravitational effects have also been studied \cite{Bluhm2004ep,Bailey2009me,Tso2011up,Kostelecky2008in,Kostelecky2010ze,Nascimento2014vva,Maluf2013nva,Maluf2014dpa,Bailey2006fd,Bluhm2007bd}, which includes gravitational waves \cite{Kostelecky2016kkn,Kostelecky2016kfm}.

Among the possibilities for an SME, the so-called {\em bumblebee model\/} is surely one of the simplest. According to it, the Lorentz symmetry breaking is induced by a {\em bumblebee field\/} $B_\mu$ as it acquires a nonzero vacuum expectation value (VEV), corresponding to the minimum of its associated potential.

In the present work, we provide a substantial contribution to the gravitational consequences of this bumblebee\linebreak model concerning its implications on the spacetime description. More specifically, it is explicitly revealed how exactly a vacuum spacetime, as described by GR, is modified by the bumblebee field when it is given by the gradient of a scalar function, besides being at its VEV. The amazing result is that any vacuum metric solution from GR will change by only a single and incredibly simple term constructed with the bumblebee field.

Although it consists of a restriction on this field, it still belongs to the whole model and, more importantly, its main advantages, just to cite a few, are that (i) finding analytical solutions becomes immensely easier, (ii) these, in turn, allow for a thorough description of any related astronomical phenomenon and (iii) it has a wide range of possible interesting applications since it deals with vacuum situations.\linebreak All this constitutes a valuable tool for seeking evidence of an underlying theory of quantum gravity, as already mentioned.

Several studies \cite{Casana:2017jkc,Ding_2021,Kanzi:2022vhp,Chen:2020qyp,Kanzi:2019gtu,Ovgun:2018ran,Yang:2018zef,Li:2020wvn,DCarvalho:2021zpf,Oliveira:2021abg,Kumar_Jha_2021} have already been made to describe the observational implications of this bumblebee\linebreak field and, thus, our results certainly provide a substantial contribution to this task.

In the next section, we describe a generic bumblebee model and its corresponding dynamical equations are displayed. Their simplified versions for vacuum are also obtained when a gradient bumblebee field at its VEV is considered.

In Sect.~\ref{Solution splitting}, with the aid of \ref{app: The Ricci tensor}, a splitting of the vacuum spacetime metric solution into the {\em background metric\/} plus the bumblebee field contributions is provided, in which the former is so called as it describes precisely the same situation in the absence of the bumblebee field. Therefore, as a method to solve for this model, it suffices to find an appropriate vacuum solution from GR to work with and a gradient bumblebee field at its VEV. Additionally, making use of the result found in \ref{app: metric determinant}, it is demonstrated that the same final form of the dynamical equations could have also been directly and easily obtained from the model action for this particular case.

As some examples, in Sect.~\ref{Some examples}, the developed method is used to determine how such a bumblebee field modifies the Schwarzschild solution and also a rotating spacetime. In the first situation, a previously published solution \cite{Casana:2017jkc} is readily obtained and it is easily generalized afterwards as a new solution. For the rotating case, a simple and new solution is also easily found. It is associated with the Kerr spacetime in the presence of the bumblebee field and it is shown in \ref{app: IRS and horizons} it keeps the same expressions for its distinguished surfaces. A rather general possibility for the bumblebee field is presented and its corresponding solution contains all the others as particular cases. Some comments are also made concerning the possibility of this type of bumblebee field deeply modifying the background spacetime structure.

Finally, in Sect.~\ref{conclusions}, we make our conclusions and final remarks.

\section{The Theoretical Model}
\label{Theoretical Model}

The bumblebee model under investigation is the same already studied in Refs.~\cite{Bluhm2004ep,Kostelecky2010ze,Maluf2013nva,Maluf2014dpa,Casana:2017jkc}, just to cite a few, and is obtained from the variational procedure applied to the following action:
\begin{align}
	S = & \int \biggl[\frac{R}{2\kappa} + \frac{\xi}{2\kappa} B^{\mu } B^{\nu} R_{\mu \nu } - \frac{1}{4}B_{\mu \nu }B^{\mu \nu } - V \nonumber \\ & + \mathcal{L}_{M}\biggr] \sqrt{-g} \, d^4x \text{,}
	\label{model action}
\end{align}
in which $B^\mu \equiv g^{\mu\nu} B_\nu$ is the contravariant version of $B_\mu$, which has the dimension of $N^{1/2}$ in SI units, $g$ is the metric determinant, $R~\equiv~g^{\mu\nu} R_{\mu\nu}$ is the {\em Ricci scalar}, $R_{\mu\nu}~\equiv~R^\alpha_{\mu\alpha\nu}$ is the {\em Ricci tensor},
\begin{equation}
	R^\alpha_{\mu\beta\nu} \equiv \partial_\beta \Gamma^\alpha_{\mu\nu} - \partial_\nu \Gamma^\alpha_{\mu\beta} + \Gamma^\alpha_{\beta\lambda}\Gamma^\lambda_{\mu\nu} - \Gamma^\alpha_{\nu\lambda}\Gamma^\lambda_{\mu\beta}
	\label{Riemann tensor definition}
\end{equation}
is the {\em Riemann tensor\/} written in terms of the {\em affine connection\/} $\Gamma^\alpha_{\mu\nu}$ given by the Christoffel symbol, $V = V\left(B^2 \pm b^2\right)$ is the potential, whose specific form will be irrelevant to our purposes and whose minimum is zero and occurs whenever its argument vanishes, $B^2 \equiv B^\mu B_\mu$, $b^{2}$ is a positive real constant with the dimension of $N$ in SI units,
\begin{equation}
	B_{\mu\nu} \equiv \partial_{\mu}B_{\nu} - \partial_{\nu}B_{\mu}
	\label{field strength}
\end{equation}
we shall call the {\em bumblebee field strength}, $\mathcal{L}_{M}$ is the Lagrangian density of any other field that may be considered, $\xi$ is a real constant with the dimension of $N^{-1}$ in SI units that controls the gravity-bumblebee field interaction, and $\kappa$ is {\em Einstein's gravitational constant\/}.

Applying the variational procedure for both the independent variations of the metric $g_{\mu\nu}$ and the bumblebee field $B_\mu$, one obtains:
\begin{subequations}\label{Field equations}
	\begin{align}
		G_{\mu\nu} & \equiv R_{\mu\nu} - \frac{1}{2}Rg_{\mu\nu} = \kappa T_{\mu\nu} \text{,} \label{Field equation - Einstein} \\
		\nabla^{\mu}B_{\mu\nu} & = 2V^{\prime}B_{\nu} - \frac{\xi}{\kappa}B^{\mu}R_{\mu\nu} \text{,} \label{Field equation - bumblebee}
	\end{align}
\end{subequations}
where $\nabla_\alpha$ indicates the covariant derivative, $G_{\mu\nu}$ is the {\em Einstein tensor\/} and
\begin{equation}
	T_{\mu\nu} = T_{\mu\nu}^{M}+T_{\mu\nu}^{B}
	\label{total energy-momentum tensor}
\end{equation}
is the total {\em energy-momentum tensor\/} split into the contributions from $\mathcal{L}_{M}$ $\left(T_{\mu\nu}^{{M}}\right)$ and from the bumblebee field $\left(T_{\mu\nu}^{B}\right)$ which, in turn, is given by
\begin{align}
	T_{\mu\nu}^{B} = & -B_{\mu\alpha}{B^{\alpha}}_{\nu} - \frac{1}{4}B_{\alpha\beta}B^{\alpha\beta}g_{\mu\nu} - Vg_{\mu\nu} + 2V^\prime B_{\mu}B_{\nu} \notag \\
	& + \frac{\xi}{2\kappa}\left[ B^{\alpha}B^{\beta}R_{\alpha\beta}g_{\mu \nu } - 2B_{\mu}B^{\alpha}R_{\alpha\nu} - 2B_{\nu}B^{\alpha}R_{\alpha\mu}\right. \notag \\
	& + \nabla_{\alpha }\nabla_{\mu }(B^{\alpha}B_{\nu}) + \nabla_{\alpha}\nabla_{\nu}(B^{\alpha}B_{\mu}) - \nabla^2 (B_{\mu}B_{\nu}) \notag \\
	& \left. - g_{\mu\nu}\nabla_{\alpha}\nabla_{\beta }(B^{\alpha}B^{\beta}) \right] \text{,}
	\label{Bumblebee EM tensor}
\end{align}
with $V^\prime$ denoting differentiation of $V$ with respect to its argument and $\nabla^2 \equiv \nabla_\alpha\nabla^\alpha$. A supposed bumblebee charge current ($J_\nu$) may be added to the right-hand side of (\ref{Field equation - bumblebee}).

Dynamical equation (\ref{Field equation - Einstein}) can be identically rewritten as:
\begin{align}
	R_{\mu \nu } = & \kappa \left( T_{\mu\nu} - \frac{1}{2}g_{\mu\nu}T\right) \notag \\
	= & \kappa \left( T_{\mu\nu}^{M} - \frac{1}{2}g_{\mu\nu}T^{M}\right) + \kappa \left( T_{\mu\nu}^{B} - \frac{1}{2}g_{\mu\nu}T^{B}\right) \text{,}
	\label{Field equation - Ricci}
\end{align}
in which we denote $T \equiv g^{\alpha\beta}T_{\alpha\beta}$ and equivalently for $T^M$ and $T^B$, which reads
\begin{equation}
	T^B = -4V + 2V^{\prime}B^2 - \frac{\xi}{\kappa} \left[ \frac{1}{2}\nabla^2 B^2 + \nabla_{\alpha}\nabla _{\beta}( B^{\alpha}B^{\beta}) \right] \text{.}
	\label{Bumblebee EM tensor trace}
\end{equation}
%with $B^2 \equiv B^\mu B_\mu$.

\subsection{The particular case of interest}
We will be particularly concerned with the special situation in which the bumblebee field is at its VEV and is also given by the gradient of a scalar field, which is a necessary and sufficient condition for a null $B_{\mu\nu}$ \cite{CourantDifIntV2}. In addition, no other sources will be taken into account, which is to say we will be dealing with vacuum. Denoting the aforementioned scalar field by $\lambda$, we have, thus:
\begin{subequations}\label{particular case}
	\begin{align}
		B^2 & = \mp b^2 \qquad \Leftrightarrow \qquad V = 0 \text{,} \quad V^\prime = 0 \text{,} \label{const bumblebee field modulus} \\
		B_\mu & = \partial_\mu \lambda \qquad \Leftrightarrow \qquad \partial_{\mu}B_{\nu} = \partial_{\nu}B_{\mu} \text{,} \label{null bumblebee field strength} \\
		\mathcal{L}_{M} & = 0 \qquad \Rightarrow \qquad T_{\mu\nu}^{M} = 0 \text{,} \label{absence of matter}
	\end{align}
\end{subequations}
and we note condition (\ref{null bumblebee field strength}) for a null bumblebee field strength is equivalent to
\begin{equation}
	\nabla_{\mu}B_{\nu} = \nabla_{\nu}B_{\mu}
	\label{null bumblebee field strength - covariant derivative}
\end{equation}
as the connection terms cancel each other out.

Dynamical equation (\ref{Field equation - bumblebee}), therefore, simplifies to:
\begin{equation}
	B^{\mu}R_{\mu\nu} = 0 \text{.}
	\label{Special field equations - bumblebee}
\end{equation}
Equation (\ref{Bumblebee EM tensor}), in turn, becomes:
\begin{align}
	T_{\mu\nu}^{B} = & \frac{\xi}{2\kappa}\Bigl[\nabla_{\alpha}\nabla_{\mu}(B^{\alpha}B_{\nu}) + \nabla_{\alpha}\nabla_{\nu}(B^{\alpha}B_{\mu}) \nonumber \\
	& - \nabla^2 (B_{\mu}B_{\nu}) - g_{\mu\nu}\nabla_{\alpha}\nabla_{\beta}(B^{\alpha}B^{\beta}) \Bigr] \text{,}
	\label{Special bumblebee EM tensor}
\end{align}
while equation (\ref{Bumblebee EM tensor trace}) is rewritten as:
\begin{equation}
	T^B = - \frac{\xi}{\kappa}\nabla_{\alpha}\nabla _{\beta}( B^{\alpha}B^{\beta}) \text{.}
	\label{Special bumblebee EM tensor trace}
\end{equation}
With (\ref{absence of matter}), (\ref{Special bumblebee EM tensor}) and (\ref{Special bumblebee EM tensor trace}), dynamical equation (\ref{Field equation - Ricci}) becomes:
\begin{align}
	R_{\mu\nu} & = \frac{\xi}{2}\left[\nabla_{\alpha}\nabla_{\mu}(B^{\alpha}B_{\nu}) + \nabla_{\alpha}\nabla_{\nu}(B^{\alpha}B_{\mu}) - \nabla^2 (B_{\mu}B_{\nu})\right] \notag \\
	& = \frac{\xi}{2}\nabla_{\alpha}\left[\nabla_{\mu}(B^{\alpha}B_{\nu}) + \nabla_{\nu}(B^{\alpha}B_{\mu}) - \nabla^\alpha (B_{\mu}B_{\nu})\right] \text{.}
	\label{Special field equations - Ricci (extended)}
\end{align}
With the aid of (\ref{null bumblebee field strength - covariant derivative}), each of the three terms inside the square brackets above can be written as:
\begin{subequations}\label{Ricci simplifications}
	\begin{align}
		\nabla_{\mu}(B^{\alpha}B_{\nu}) & = B^{\alpha}\nabla_{\mu}B_{\nu} + B_{\nu}\nabla_{\mu}B^{\alpha} = B^{\alpha}\nabla_{\mu}B_{\nu} + B_{\nu}\nabla^{\alpha}B_{\mu} \text{,} \label{Ricci simplification 01} \\
		\nabla_{\nu}(B^{\alpha}B_{\mu}) & = B^{\alpha}\nabla_{\nu}B_{\mu} + B_{\mu}\nabla_{\nu}B^{\alpha} = B^{\alpha}\nabla_{\mu}B_{\nu} + B_{\mu}\nabla^{\alpha}B_{\nu} \text{,} \label{Ricci simplification 02} \\
		\nabla^\alpha (B_{\mu}B_{\nu}) & = B_{\nu}\nabla^{\alpha}B_{\mu} + B_{\mu}\nabla^{\alpha}B_{\nu} \text{,} \label{Ricci simplification 03}
	\end{align}
\end{subequations}
and (\ref{Special field equations - Ricci (extended)}) simplifies to:
\begin{equation}
	R_{\mu\nu} = \xi\nabla_{\alpha}(B^{\alpha}\nabla_\mu B_\nu ) \text{.}
	\label{Special field equations - Ricci (simplified)}
\end{equation}

Contraction of this equation with $B^\mu$ gives:
\begin{equation}
	B^\mu R_{\mu\nu} = \xi\left[\nabla_{\alpha}(B^{\alpha}B^\mu\nabla_\mu B_\nu ) - (\nabla_\mu B_\nu)B^{\alpha}\nabla_{\alpha}B^\mu \right] \text{.}
	\label{Special field equations - Ricci (contracted)}
\end{equation}
However, Eq.~(\ref{const bumblebee field modulus}) for the bumblebee field modulus along with Eq.~(\ref{null bumblebee field strength - covariant derivative}) imply:
\begin{equation}
	B^\mu \nabla_\nu B_\mu = B^\mu \nabla_\mu B_\nu = 0
	\label{Differential consequences for special B}
\end{equation}
and, thus, both terms inside the square brackets of (\ref{Special field equations - Ricci (contracted)}) are zero. Consequently, (\ref{Special field equations - bumblebee}) is automatically satisfied by (\ref{Special field equations - Ricci (simplified)}), which is, then, the only dynamical equation we need to solve in this case.

Although dynamical equations (\ref{Field equations}) have already been considerably simplified in this particular scenario, an even further simplification will be performed in the next section, where the spacetime metric solution will be given precisely in terms of its GR corresponding plus contributions from the bumblebee field.

\section{Exact effect of the bumblebee field on the spacetime}
\label{Solution splitting}

In this section, we shall present how exactly this kind of bumblebee field contributes to the metric solution. For that purpose, let us consider the following tensor:
\begin{subequations}\label{background metric definitions}
	\begin{equation}
		\tilde{g}_{\mu\nu} \equiv g_{\mu\nu} - \frac{\xi}{1 + \xi B^2}B_\mu B_\nu \text{,}
		\label{background metric (xi explicit)}
	\end{equation}
	whose {\em inverse\/} is:
	\begin{equation}
		\tilde{g}^{\mu\nu} \equiv g^{\mu\nu} + \xi B^\mu B^\nu \text{.} \label{inverse background metric (xi explicit)}
	\end{equation}
\end{subequations}

As a simple consequence, condition (\ref{const bumblebee field modulus}) for the bumblebee field to be at its VEV can also be equivalently rewritten as:
\begin{equation}
	\tilde{B}^2 \equiv \tilde{g}^{\mu\nu}B_\mu B_\nu = B^2\left(1 + \xi B^2\right) = \mp b^2\left(1 \mp \xi b^2\right) \text{,}
	\label{bumblebee modulus via background metric}
\end{equation}
which can be regarded as its modulus according to $\tilde{g}_{\mu\nu}$ and it is really noteworthy it is still a constant. Even though this result is actually useful to our approach, the great benefit of the above definition is better appreciated when equation (\ref{Special field equations - Ricci (simplified)}) is also written in terms of it.

In \ref{app: The Ricci tensor}, it is shown that substituting $g_{\mu\nu}$ and $g^{\mu\nu}$ from Eqs.~(\ref{background metric definitions}) in the Ricci tensor results in
\begin{equation}
	R_{\mu\nu} = \tilde{R}_{\mu\nu} + \xi\nabla_{\alpha}(B^{\alpha}\nabla_\mu B_\nu ) \text{,}
	\label{background Ricci tensor (xi explicit)}
\end{equation}
with $\tilde{R}_{\mu\nu}$ being the Ricci tensor completely written in terms of $\tilde{g}_{\mu\nu}$ instead of $g_{\mu\nu}$.

Making use of this result in Eq.~(\ref{Special field equations - Ricci (simplified)}), it becomes simply:
\begin{equation}
	\tilde{R}_{\mu\nu} = 0 \text{,}
	\label{Special field equations - background Ricci}
\end{equation}
which is exactly Einstein's equation of GR in vacuum for $\tilde{g}_{\mu\nu}$.

Therefore, as a very simple method to find solutions to the model described by action (\ref{model action}) in vacuum, one only needs to take a $\tilde{g}_{\mu\nu}$ that describes a vacuum spacetime from GR (\ref{Special field equations - background Ricci}) along with a bumblebee field that satisfies both
\begin{subequations}\label{bumblebee field constraints}
	\begin{align}
		\tilde{g}^{\mu\nu}B_\mu B_\nu & = \tilde{B}^2 \text{,} \label{constant bumblebee modulus via background metric} \\
		\partial_\mu B_\nu & = \partial_\nu B_\mu \text{.} \label{null bumblebee field strength (repeated)}
	\end{align}
\end{subequations}
Then, from (\ref{background metric (xi explicit)}), the metric given by
\begin{equation}
	g_{\mu\nu} = \tilde{g}_{\mu\nu} + \frac{\xi}{1 + \xi B^2}B_\mu B_\nu \text{,}
	\label{metric in terms of the background metric}
\end{equation}
with $B^2$ given by (\ref{const bumblebee field modulus}), will automatically solve the corresponding system of dynamical equations.

It is actually remarkable how this explicitly reveals the exact effect on the spacetime geometry caused by the presence of such kind of bumblebee field, as the metric would be given only by $\tilde{g}_{\mu\nu}$ in the absence of it. For that precise reason, we shall call it the {\em background metric\/} from now on.

Before we finish this section, however, we draw attention to the fact that this simplification could also have been easily and directly obtained from action (\ref{model action}). As conditions (\ref{const bumblebee field modulus}) and (\ref{null bumblebee field strength}) consist of holonomic constraints, both of them, along with (\ref{absence of matter}), simplify the model action as
\begin{subequations}\label{simplified action}
	\begin{align}
		S & = \int \frac{1}{2\kappa} (R + \xi B^{\mu} B^{\nu} R_{\mu\nu}) \sqrt{-g} \, d^4x \label{simplified model action} \\
		& = \int \frac{1}{2\kappa} (g^{\mu\nu} + \xi B^{\mu} B^{\nu}) R_{\mu\nu} \sqrt{-g} \, d^4x \notag  \\
		& = \int \frac{1}{2\kappa} \tilde{g}^{\mu\nu}\left[\tilde{R}_{\mu\nu} + \xi\nabla_{\alpha}(B^{\alpha}\nabla_\mu B_\nu )\right]\sqrt{-g} \, d^4x \text{,} \label{simplified background model action (preliminary)}
	\end{align}
\end{subequations}
in which equations (\ref{inverse background metric (xi explicit)}) and (\ref{background Ricci tensor (xi explicit)}) have been used to obtain (\ref{simplified background model action (preliminary)}). In this expression, we have:
\begin{align}
	\tilde{g}^{\mu\nu}\nabla_{\alpha}(B^{\alpha}\nabla_\mu B_\nu ) & = (g^{\mu\nu} + \xi B^{\mu} B^{\nu})\nabla_{\alpha}(B^{\alpha}\nabla_\mu B_\nu ) \nonumber \\
	& = \nabla_{\alpha}(B^{\alpha}\nabla_\mu B^\mu ) + \xi B^{\mu}B^{\nu}\nabla_{\alpha}(B^{\alpha}\nabla_\mu B_\nu ) \text{.}
\end{align}
However, we have already seen from equations (\ref{Special field equations - Ricci (simplified)})--(\ref{Differential consequences for special B}) that the second term on the right-hand side is zero. The first term, in turn, will give rise to a total divergence when multiplied by $\sqrt{-g}$ and will result in a surface term in the action, which will not contribute to the dynamical equations since the variational procedure does not consider variations over that surface. Then, apart from this term, action (\ref{simplified background model action (preliminary)}) becomes:
\begin{align}
	S & = \int \frac{1}{2\kappa} \tilde{R}\sqrt{-g} \, d^4x \text{,}
	\label{simplified background model action (metric determinant)}
\end{align}
in which $\tilde{R} \equiv \tilde{g}^{\mu\nu}\tilde{R}_{\mu\nu}$ is the Ricci scalar for the background metric.

In \ref{app: metric determinant}, it is shown we may write $g$ in terms of $\tilde{g} \equiv \det[\tilde{g}_{\mu\nu}]$ as
\begin{equation}
	g = \tilde{g}(1+\xi B^2 )
	\label{background metric determinant}
\end{equation}
and, hence, the action is rewritten as
\begin{align}
	S & = \int \frac{1}{2\tilde{\kappa}} \tilde{R}\sqrt{-\tilde{g}} \, d^4x \text{,}
	\label{simplified background model action (final form)}
\end{align}
with $\tilde{\kappa} \equiv \kappa/\sqrt{1+\xi B^2}$.

Written this way, it becomes very clear why the variational procedure applied to both independent variations of $g_{\mu\nu}$ and $B_\mu$ turns out to provide only equation (\ref{Special field equations - background Ricci}) to solve.

Equations (\ref{Special field equations - background Ricci})--(\ref{metric in terms of the background metric}) concisely describe this particular model and considerably simplify its mathematical treatment. In particular, they provide a truly advantageous method when already known vacuum spacetimes from GR are considered, as they allow us to skip (\ref{Special field equations - background Ricci}) and go straight to (\ref{bumblebee field constraints}) to find a solution.

In the next section, we will see some examples of how we can benefit from this possibility to easily obtain new solutions to this model.

\section{Examples of solutions obtained via the developed method}
\label{Some examples}

Before implementing the method, and especially in view of some peculiar results obtained in the following examples, it is opportune to make some preliminary remarks about the expectations one may have concerning the possible outcomes as well as what should be the best perspective when dealing with this scenario.

As it follows from (\ref{metric in terms of the background metric}), the solution can be so different from the background metric that they may not even share the same spacetime properties. More specifically, it is possible to have a background metric consistent with some sorts of symmetries, e.g., spherical or axial, which will not be followed by the resulting metric.

Of course, if the purpose is just to find a new solution irrespective of the outcome, then this is not actually an issue. In fact, any result is worthy of attention, especially when investigating the possible consequences a theoretical model can bring. Nevertheless, sometimes one may be particularly interested in some specific spacetime configurations and compatibility with (\ref{bumblebee field constraints}) may impose serious restrictions on that.

When the solution is supposed to have some sort of properties, it is quite natural to consider a  background metric in accordance with them. However, in some cases, the corresponding possibilities for a compatible bumblebee field may happen to be so restrictive that it becomes really impossible for the resulting metric to meet with the same features.

If it turns out to be necessary to solve Eq.~(\ref{Special field equations - background Ricci}) uniquely to comply with some specific spacetime attributes, then the method would be pointless; it would be simpler to solve (\ref{Special field equations - Ricci (simplified)}) directly for the metric indeed.

On the other hand, with regard to the possibility of solving either of these equations with that single purpose, it is not clear, however, if this particular model can account for every kind of geometry. As conditions (\ref{bumblebee field constraints}) are somewhat restrictive, perhaps there will not be always a corresponding background metric to any type of metric solution one may be interested in. It is not unlikely that some cases will demand at least one of these conditions to be relaxed and, consequently, a direct approach to (\ref{Field equations}) in this broader context becomes necessary as the method would not apply then.

Therefore, instead of seeking a potentially nonexistent fine-tuned pair of $\tilde{g}_{\mu\nu}$ and $B_\mu$ to meet with some desired features, it would be better to forsake that expectancy and simply consider how the presence of such a bumblebee field changes the spacetime, regardless of whether its properties are kept or not. This will be the perspective we shall adopt from now on.

Despite all this, it is indeed undeniable that finding solutions is considerably easier through this procedure and if they do not match the expectations, it does not compromise the method at all, as it is rather related to the model itself when this particular type of bumblebee field is considered. Besides, any result provided must be regarded as a possible solution to (\ref{Field equations}) and can reveal very important aspects of the theory, especially the most peculiar ones it may admit.

Having said that, it will be investigated next how a gradient bumblebee field at its VEV changes, first, a static and spherically symmetric vacuum spacetime and, afterwards, the more general stationary and rotating one. In both situations, the bumblebee field can make the resulting spacetime deviate from those symmetries. Nevertheless, for the first case, it is still possible to recover the spherical symmetry as a particular solution. For the rotating spacetime, on the other hand, there will be no such possibility, i.e., the metric and the background metric will never share the same spacetime properties.

\subsection{Static and spherically symmetric spacetime}
\label{sec: Schwarzrschild-like solution}

As a first and simple example, let us consider how the presence of a bumblebee field compatible with (\ref{bumblebee field constraints}) modifies the Schwarzschild spacetime, which is the most general static and spherically symmetric vacuum solution from GR \cite{Wald}.

In terms of spherical spacetime coordinates, given by $x^\mu = (t, r, \theta, \varphi)$, the background metric then reads
\begin{equation}
	\tilde{g}_{\mu\nu} = \operatorname{diag}\left[-\left( 1 - \frac{2M}{r} \right), \frac{1}{1 - \frac{2M}{r}}, r^2, r^2\sin^2\theta \right]
	\label{Schwarzschild background metric}
\end{equation}
with $M$ being an arbitrary constant with the dimension of length.

Before taking the most general compatible bumblebee field, let us consider first, just for simplicity:
\begin{equation}
	B_{\mu} = \left( 0,b_{1}(r),0,0\right) \text{,}
	\label{Schwarzschild-like generic bumblebee}
\end{equation}
which naturally satisfies (\ref{null bumblebee field strength (repeated)}) and is also static and spherically symmetric. Imposing the remaining condition (\ref{constant bumblebee modulus via background metric}) readily gives:
\begin{subequations}\label{Schwarzschild-like solution}
	\begin{equation}
		B_\mu = \pm \frac{\tilde{B}}{\sqrt{1 - \frac{2M}{r}}} \delta^1_\mu \text{,}
		\label{Schwarzschild-like specific bumblebee}
	\end{equation}
	with $\tilde{B} \equiv \sqrt{\tilde{B}^2}$ and the plus sign in (\ref{const bumblebee field modulus}) is considered, as this field is spacelike.
	
	Therefore, according to (\ref{metric in terms of the background metric}), and just as simple as that, we are brought to the solution
	\begin{equation}
		g_{\mu\nu} = \operatorname{diag}\left[-\left( 1 - \frac{2M}{r} \right), \frac{1 + \ell}{1 - \frac{2M}{r}}, r^2, r^2\sin^2\theta \right]
		\label{Schwarzschild-like published solution}
	\end{equation}
\end{subequations}
which is precisely the same found in Ref.~\cite{Casana:2017jkc} written in terms of the dimensionless constant $\ell \equiv \xi b^2$, but obtained much more easily.

In this particular case, the resulting spacetime is still static and spherically symmetric, but this is not exclusive to the particular form of Eq.~(\ref{Schwarzschild-like specific bumblebee}). That would also be the case if we had considered the following more general bumblebee field instead:
\begin{subequations}\label{Schwarzschild-like generic bumblebee 02 complete}
	\begin{equation}
		B_{\mu} = \left( b_0,b_{1}(r),0,0\right) \text{,}
		\label{Schwarzschild-like generic bumblebee 02}
	\end{equation}
	for an arbitrary constant $b_0$, which is still in accordance with the spacetime symmetries.
	
	Although it may not seem to provide a spherically symmetric spacetime, for it will give rise to a nonzero $g_{01}$ as a consequence of (\ref{metric in terms of the background metric}), we will see it is possible to bring the metric to the same form of (\ref{Schwarzschild-like published solution}) by performing a suitable coordinate transformation.
	
	Condition (\ref{null bumblebee field strength (repeated)}) is already satisfied and (\ref{constant bumblebee modulus via background metric}) imposes
	\begin{equation}
		b_{1}(r) = \pm \frac{\sqrt{\tilde{B}^2\left( 1 - \frac{2M}{r} \right) + b_0^2}}{1 - \frac{2M}{r}} \text{.}
		\label{Schwarzschild-like specific bumblebee 02}
	\end{equation}
\end{subequations}
\begin{subequations}\label{Schwarzschild-like general metric}
	As a result, the metric acquires the following nonzero components:
	\begin{align}
		g_{00} & = -\left( 1 - \frac{2M}{r} \right) + \frac{\xi b_0^2}{1 + \xi B^2} \notag \\ & = -\left[ \frac{1-\xi\left( b_0^2 - B^2 \right)}{1 + \xi B^2} - \frac{2M}{r} \right] \text{,}  \label{Schwarzschild-like general g00} \\
		g_{01} & = \pm \frac{\xi b_0}{1 + \xi B^2}\cdot\frac{\sqrt{\tilde{B}^2\left( 1 - \frac{2M}{r} \right) + b_0^2}}{1 - \frac{2M}{r}} \text{,} \label{Schwarzschild-like general g01} \\
		g_{11} & = \frac{1}{1 - \frac{2M}{r}} + \frac{\xi}{1 + \xi B^2}\cdot\frac{\left[\tilde{B}^2\left( 1 - \frac{2M}{r} \right) + b_0^2\right]}{\left( 1 - \frac{2M}{r} \right)^2} \text{,} \label{Schwarzschild-like general g11} \\
		g_{22} & = r^2 \text{,} \label{Schwarzschild-like general g22} \\
		g_{33} & = r^2 \sin^2\theta \text{.} \label{Schwarzschild-like general g33}
	\end{align}
\end{subequations}
The sign in $g_{01}$ will be the same as in (\ref{Schwarzschild-like specific bumblebee 02}).

Even though this is already a new solution to this model, we may still remove that unwanted $g_{01}$ by just performing the transformation:
\begin{equation}
	t = \Bar{t} + T(r) \quad \left| \quad \frac{dT}{dr} = -\frac{g_{01}}{g_{00}} \right. \text{.}
	\label{coord transf for t}
\end{equation}
Since both $g_{01}$ and $g_{00}$ are functions of $r$ only, one can integrate for $T(r)$, but we shall not do it as this is not necessary for our purposes.

Besides eliminating $g_{01}$ as a result of that transformation, the metric will also change only its component $g_{11}$, which now reads
\begin{equation}
	g_{11} = \frac{1 + \xi B^2}{\frac{1-\xi\left( b_0^2 - B^2 \right)}{1 + \xi B^2} - \frac{2M}{r}} \text{.}
	\label{Schwarzschild-like general g11 transformed}
\end{equation}
The metric is now diagonal, but a further simplifying manipulation can still be done.

Writing the arbitrary constant $M$ as:
\begin{equation}
	M \equiv M^\prime \left[ \frac{1-\xi\left( b_0^2 - B^2 \right)}{1 + \xi B^2} \right] \text{,}
	\label{redefinition of M}
\end{equation}
the metric components $g_{00}$ and $g_{11}$ become
\begin{subequations}\label{Schwarzschild-like g00 and g01 (M prime)}
	\begin{align}
		g_{00} & = -\frac{\left[ 1-\xi\left( b_0^2 - B^2 \right) \right]}{1 + \xi B^2} \left( 1 - \frac{2M^\prime}{r} \right) \text{,} \label{Schwarzschild-like g00 (M prime)} \\
		g_{11} & = \frac{\left(1 + \xi B^2\right)^2}{1-\xi\left( b_0^2 - B^2 \right)}\cdot\frac{1}{\left( 1 - \frac{2M^\prime}{r} \right)} \label{Schwarzschild-like g11 (M prime)} \text{.}
	\end{align}
\end{subequations}
The constant fraction on (\ref{Schwarzschild-like g00 (M prime)}) can be incorporated to $\Bar{t}$, which actually consists of making the new transformation:
\begin{equation}
	\Bar{t} = \tilde{t} \sqrt{\frac{1 + \xi B^2}{1-\xi\left( b_0^2 - B^2 \right)}} \text{,}
	\label{coord transf for t bar}
\end{equation}
and results in
\begin{equation}
	g_{00} = -\left( 1 - \frac{2M^\prime}{r} \right) \text{.}
	\label{Schwarzschild-like g00 (M prime) 2nd transf}
\end{equation}

The bumblebee field components, in turn, after the successive transformations (\ref{coord transf for t}) and (\ref{coord transf for t bar}), becomes
\begin{subequations}\label{Schwarzschild-like transformed bumblebee field}
	\begin{align}
		B_0 & = b_0 \sqrt{\frac{1 + \xi B^2}{1-\xi\left( b_0^2 - B^2 \right)}} \text{,} \label{Schwarzschild-like transformed B0} \\
		B_1 & = \pm \frac{\sqrt{\left( 1 + \xi B^2 \right)B^2 \left( 1 - \frac{2M}{r} \right) + b_0^2}}{\frac{1-\xi\left( b_0^2 - B^2 \right)}{1 + \xi B^2} - \frac{2M}{r}} \notag \\
		& = \pm \frac{\sqrt{\left[ 1 + \xi( B^2 + B_0^2) \right]\left[\left( 1 - \frac{2M^\prime}{r} \right)B^2 + B_0^2\right]}}{1 - \frac{2M^\prime}{r}} \text{.} \label{Schwarzschild-like transformed B1}
	\end{align}
\end{subequations}
To obtain this last expression for $B_1$, equation (\ref{redefinition of M}) was used together with the expression for $b_0$ in terms of $B_0$ that comes from (\ref{Schwarzschild-like transformed B0}):
\begin{equation}
	b_0 = B_0 \sqrt{\frac{1 + \xi B^2}{1 + \xi\left( B_0^2 + B^2 \right)}} \text{.}
	\label{b0 in terms of B0}
\end{equation}

Thus, after taking the above expression also in (\ref{Schwarzschild-like g11 (M prime)}), the solution finally reads
\begin{subequations}\label{{Schwarzschild-like general solution}}
	\begin{align}
		B_\mu & = \left( B_0, \pm \frac{\sqrt{( 1 + \ell^\prime )\left[\left( 1 - \frac{2M^\prime}{r} \right)B^2 + B_0^2\right]}}{1 - \frac{2M^\prime}{r}}, 0, 0 \right) \text{,} \label{Schwarzschild-like general bumblebee field} \\
		g_{\mu\nu} & = \operatorname{diag}\left[-\left( 1 - \frac{2M^\prime}{r} \right), \frac{1 + \ell^\prime}{1 - \frac{2M^\prime}{r}}, r^2, r^2\sin^2\theta \right] \text{,}
		\label{Schwarzschild-like general metric solution}
	\end{align}
	with the dimensionless constant $\ell^\prime$ given by
	\begin{equation}
		\ell^\prime \equiv \xi\left( B^2 + B_0^2 \right) \text{.}
		\label{ell prime definition}
	\end{equation}
\end{subequations}

We point out that $B_0$ remains an arbitrary constant and this new solution naturally recovers (\ref{Schwarzschild-like solution}) just by making $B_0 = 0$, along with $B^2 = b^2$, as expected.

Concerning the metric, since both $M$ and $M^\prime$ are arbitrary constants, this spacetime solution only differs from (\ref{Schwarzschild-like published solution}) on $\ell^\prime$, both being qualitatively equal though. As a consequence, the same upper-bound estimate of order $10^{-19}$ to $|\ell|$ found in Ref.~\cite{Casana:2017jkc} now applies to $|\ell^\prime|$:
\begin{equation}
	|\ell^\prime| = |\xi|\left| B^2 + B_0^2 \right| \lesssim 10^{-19} \text{.}
	\label{constraint on ell prime}
\end{equation}

Due to the arbitrariness of $B_0$, this condition now becomes meaningless with regard to specifying or constraining the parameters of the theory. In fact, even in the previous case for which $B_0$ is zero this is also true, as the upper-bound would be imposed to $|\ell| = |\xi| b^2$, leaving both parameters $\xi$ and $b^2$ still unconstrained individually.

Besides, since the bumblebee field now may be either spacelike or timelike, both signs in (\ref{const bumblebee field modulus}) are possible. For $B^2 = b^2$, any nonzero value for $B_0$ makes the condition for $|\ell|$ even more restrictive. For $B^2 = -b^2$, in turn, $|\ell|$ loses any restriction at all. 

With respect to the observational consequences, on the other hand, condition (\ref{constraint on ell prime}) is actually very meaningful, as it imposes they must be very constrained as compared to the Schwarzschild solution.

Since the intent is just to present an example of an application of the method described in the last section and, in addition, the solution obtained is also qualitatively equal to the already studied solution found in Ref.~\cite{Casana:2017jkc}, no further development will be made concerning this result.

Before moving forward to the next example and just for completion, it is worth considering a last, and supposedly most general, possibility for the bumblebee field compatible with (\ref{bumblebee field constraints}) given by:
\begin{subequations}\label{Schwarzschild-like generic bumblebee B2 and B3 nonzero complete}
	\begin{equation}
		B_{\mu} = \left( B_0,B_{1}(r),B_{2}(\theta),B_3\right) \text{.}
		\label{Schwarzschild-like generic bumblebee B2 and B3 nonzero}
	\end{equation}
	Components $B_0$ and $B_3$ are arbitrary constants and the others are expressed as:
	\begin{align}
		B_{1}(r) & = \pm \frac{\sqrt{\tilde{B}^2\left(1 - \frac{C}{r^2}\right)\left(1 - \frac{2M}{r}\right) + B_0^2}}{1 - \frac{2M}{r}} \text{,} \label{Schwarzschild-like generic bumblebee B2 and B3 nonzero B1} \\
		B_{2}(\theta) & = \pm \sqrt{\tilde{B}^2 C - \frac{B_3^2}{\sin^2\theta}} \text{,} \label{Schwarzschild-like generic bumblebee B2 and B3 nonzero B2}
	\end{align}
\end{subequations}
for an arbitrary constant $C$ with the dimension of length squared and unrelated sign possibilities. The previous example consists of the particular case in which both $B_3$ and $C$ are zero.

As a nonvanishing $B_3$ makes component $B_2$ ill-defined, we shall set the former to zero and the field is rewritten as:
\begin{subequations}\label{Schwarzschild-like generic bumblebee B2 nonzero complete}
	\begin{gather}
		B_{\mu} = \left( B_0,B_{1}(r),B_2,0\right) \text{,} \label{Schwarzschild-like generic bumblebee B2 nonzero} \\
		B_{1}(r) = \pm \frac{\sqrt{\left(\tilde{B}^2 - \frac{B_2^2}{r^2}\right)\left(1 - \frac{2M}{r}\right) + B_0^2}}{1 - \frac{2M}{r}} \text{,} \label{Schwarzschild-like generic bumblebee B2 nonzero B1}
	\end{gather}
\end{subequations}
for an arbitrary constant component $B_2$.

Regardless of any mathematical issue a nonzero $B_2$ may bring to the solution, the resulting metric will have only components $g_{\mu 3}$ ($\mu \neq 3$) equal to zero and will clearly deviate from the spherical symmetry. In this case, there is no coordinate transformation that could even lead it to the diagonal form, making this an example of a solution that does not follow the same spacetime aspects of the background metric. In other words, the presence of such a bumblebee field breaks the spherical symmetry the spacetime would have in the absence of it.

Nevertheless, as we have previously seen, it is just a matter of setting $B_2$ to zero to keep the same features also shared by $\tilde{g}_{\mu\nu}$. In the following example, however, such compatibility will never be possible.

Although the solution that results from (\ref{Schwarzschild-like generic bumblebee B2 nonzero complete}) is surely interesting and worthwhile studying, we do not intend to explore it further at the moment, as the focus is on the application of the method rather than deeply investigating the solutions it provides.

\subsection{Stationary and rotating spacetime}
\label{sec: Kerr-like solution}

As a second example, we will consider the effects a gradient bumblebee field at its VEV may cause on a rotating vacuum spacetime from GR. However, instead of taking the Kerr metric at first, we will consider a rather more general one that can be obtained following the lines of Ref.~\cite{Klotz1982} and recovers it as a particular case. The background metric will be given by
\begin{subequations}\label{rotating general background metric complete}
	\begin{equation}
		\left[\tilde{g}_{\mu\nu}\right] = \begin{bmatrix}
			-\gamma & 0 & 0 & -\frac{q(1 - \gamma)}{a} \\
			0 & \frac{A(p - q)}{\Delta} & 0 & 0 \\
			0 & 0 & A(p - q) & 0 \\
			-\frac{q(1 - \gamma)}{a} & 0 & 0 & \frac{q[p + (1 - \gamma)q]}{a^2} 
		\end{bmatrix}
		\label{rotating general background metric}
	\end{equation}
	in which spherical spacetime coordinates are being used, $A$ is dimensionless, $a$ has the dimension of length and both of them are arbitrary constants. The remaining symbols indicate functions whose arguments were omitted for simplicity's sake and are given by
	\begin{gather}
		p = p(r) = r^2 + N \text{,} \label{general p} \\
		q = q(\theta) = N\sin^2{\theta} \text{,} \label{general q}
	\end{gather}
	\begin{gather}
		\gamma = \gamma(r,\theta) = 1 - \frac{2Mr}{\rho^2} \text{,} \label{general gamma} \\
		\Delta = \Delta(r) = r^2 + N - 2Mr \text{,} \label{general Delta} \\
		\rho^2 = \rho^2(r,\theta) = r^2 + N\cos^2{\theta} \text{,} \label{general rho2}
	\end{gather}
\end{subequations}
in terms of the arbitrary constants $M$ and $N$, with the dimensions of length and length squared, respectively. The Kerr metric corresponds to the particular case in which $A = 1$ and $N = a^2$.

For the bumblebee field, a rather general possibility compatible with (\ref{bumblebee field constraints}) is given by
\begin{subequations}\label{compatible rotating bumblebee field complete}
	\begin{equation}
		B_\mu = (B_0, B_1(r), B_2(\theta), B_3) \text{,}
		\label{compatible rotating bumblebee field}
	\end{equation}
	with $B_0$ and $B_3$ being both arbitrary constants whereas $B_1(r)$ and $B_2(\theta)$ are expressed as
	\begin{align}
		B_1(r) & = \pm \sqrt{\frac{A}{\Delta}\left[\tilde{B}^2 (p - C) - 2 a B_0 B_3 + \frac{(B_0 p + a B_3 )^2}{\Delta}\right]} \\
		B_2(\theta) & = \pm \sqrt{A\left[\tilde{B}^2(C - q) - B_0^2q - \frac{a^2 B_3^2}{q} \right]} \text{.} \label{general rotating B2}
	\end{align}
\end{subequations}
The sign possibilities are unrelated and $C$ is an arbitrary constant with the dimension of length squared.

We point out that any proposal with either $B_1$ or $B_2$ as a constant fails to meet with the required conditions, so the arbitrariness necessarily concerns only the other components indeed.

The second to last case from the previous section, given by the Schwarzschild spacetime modified by (\ref{Schwarzschild-like generic bumblebee B2 and B3 nonzero complete}), is recovered by making $A = 1$, $N = a^2$ and, then, $a = 0$.

Once again, to avoid an ill-defined bumblebee field, we shall set $B_3$ to zero and the resulting metric solution, then, will have only components $g_{13}$ and $g_{23}$ equal to zero. As interesting as it can be, exploring such a complex spacetime is not our point and may be the subject of a future publication. Thus, for the sake of simplicity, we would rather concentrate on the least modifications the bumblebee field may introduce, which consists of setting both $B_0$ and $B_3$ to zero. As it will be spacelike in this case, it must be $B^2 = b^2$ and, in terms of $\Delta$ and $\rho^2$, it provides:
\begin{subequations}\label{particular rotating bumblebee field complete}
	\begin{equation}
		B_\mu = \left(0, B_1(r), B_2(\theta), 0\right) \text{,}
		\label{particular rotating bumblebee field}
	\end{equation}
	with
	\begin{align}
		B_1(r) & = \pm b\sqrt{\frac{A(1 + \ell)(r^2 + N - C)}{\Delta}} \text{,} \label{particular rotating bumblebee field B1} \\
		B_2(\theta) & = \pm b\sqrt{A(1 + \ell)(C - N\sin^2{\theta})} \label{particular rotating bumblebee field B2}
	\end{align}
\end{subequations}
and the spacetime metric, in turn, will have the following nonzero components:
\begin{subequations}\label{particular rotating metric complete}
	\begin{align}
		g_{00} & = -\left(1 - \frac{2Mr}{\rho^2}\right) \text{,} \\
		g_{03} & = -\frac{2MNr\sin^2\theta}{a\rho^2} \text{,} \\
		g_{11} & = \frac{A\left[\rho^2 + \ell r^2 - \ell(C - N)\right]}{\Delta} \text{,} \\
		g_{12} & = \pm \ell A \sqrt{\frac{(r^2 + N - C)(C - N\sin^2\theta)}{\Delta}} \text{,} \label{particular rotating g12} \\
		g_{22} & = A\left[\rho^2 + N\ell\cos^2\theta + \ell(C - N)\right] \text{,} \\
		g_{33} & = \frac{N}{a^2}\left(r^2 + N + \frac{2MNr\sin^2\theta}{\rho^2}\right)\sin^2\theta \text{.} \label{particular rotating g33}
	\end{align}
\end{subequations}
The sign in $g_{12}$ will be plus or minus whenever both sign choices in (\ref{particular rotating bumblebee field B1}) and (\ref{particular rotating bumblebee field B2}) are the same or not, respectively.

As both $B_1$ and $B_2$ must be different from zero, so must be $g_{12}$ as well, but we stress once again this only happens because of the particular type of bumblebee field the method requires. Of course, one could perfectly well try a different expression for it in the hope the metric would keep the same form as the background metric, but a direct approach to (\ref{Field equations}) would be necessary in that case. Nonetheless, it is actually remarkable how easily a solution was found employing the method we developed, as peculiar it is.

If we wish to recover the Kerr solution when the bumblebee field contributions are removed, we should set $A = 1$ and $N = a^2$. If we make $a = 0$ yet, then we recover the last possibility from the previous section, concerning Schwarzschild spacetime modified by (\ref{Schwarzschild-like generic bumblebee B2 nonzero complete}), with $B_0 = 0$. Making $C = 0$ afterwards, it gives solution (\ref{Schwarzschild-like solution}).

A simplifying choice consists of setting $C = N$ and, based on the last paragraph, if we also set $A = 1$ and $N = a^2$, both the Kerr and the Schwarzschild-like solution given by (\ref{Schwarzschild-like solution}) are easily recovered by simply making either $b = 0$ or $a = 0$, respectively.

With that particular choice for $A$, $C$ and $N$, the solution reads:
\begin{equation}
	B_\mu = \left(0, \pm b\sqrt{1 + \ell}\frac{r}{\sqrt{\Delta}}, \pm ab\sqrt{1 + \ell}\cos\theta, 0\right)
	\label{particular Kerr-like bumblebee field}
\end{equation}
and, for the nonzero metric components:
\begin{subequations}\label{Kerr-like metric solution}
	\begin{align}
		g_{00} & = -\left(1 - \frac{2Mr}{\rho^2}\right) \text{,} \label{Kerr-like g00} \\
		g_{03} & = -\frac{2Mar\sin^2\theta}{\rho^2} \text{,} \\
		g_{11} & = \frac{\rho^2 + \ell r^2}{\Delta} \text{,}
	\end{align}
	\begin{align}
		g_{12} & = \pm \frac{\ell a r\cos\theta}{\sqrt{\Delta}} \text{,} \label{Kerr-like g12}\\
		g_{22} & = \rho^2 + \ell a^2\cos^2\theta \text{,} \\
		g_{33} & = \left(r^2 + a^2 + \frac{2Ma^2r\sin^2\theta}{\rho^2}\right)\sin^2\theta \text{,}
	\end{align}
	while (\ref{general Delta}) and (\ref{general rho2}) becomes:
	\begin{align}
		\Delta & = r^2 + a^2 - 2Mr \text{,} \label{Kerr-like Delta} \\
		\rho^2 & = r^2 + a^2\cos^2{\theta} \text{.} \label{Kerr-like rho2}
	\end{align}
\end{subequations}

Despite being a particular case of (\ref{particular rotating bumblebee field complete}) and (\ref{particular rotating metric complete}), this new solution is really worthwhile studying, especially because of its simplicity to describe a stationary and axially symmetric rotating spacetime in this bumblebee-gravity scenario.

Its many aspects, including the consequences of that non\-zero $g_{12}$, shall be the subject of a future publication but, before concluding, we can already anticipate some of its most characteristic features.

As it is shown in \ref{app: IRS and horizons}, despite the differences from the Kerr metric, its distinguished surfaces are still given by the same expressions, which are:
\begin{equation}
	r_\infty^\pm = M \pm \sqrt{M^2 - a^2\cos^2\theta} \text{,}
	\label{infinite redshift surfaces}
\end{equation}
for the infinite redshift surfaces, and
\begin{equation}
	r_\pm = M \pm \sqrt{M^2 - a^2} \text{,}
	\label{Kerr-like horizon}
\end{equation}
for the horizons.

It is interesting to note that in Ref.~\cite{Ding_2021} the authors sought slowly rotating black hole solutions in this context. The one they obtained for case A of their treatment is perfectly consistent with the method we developed which, in turn, also explains why their solution for case B still demands extra conditions. As it lacks the corresponding version of (\ref{Kerr-like g12}), their solution to this case will only be valid if $\ell a$ is negligible as well. Further analysis of the slow rotation solution for case A, especially with respect to its observational consequences, can be found in that same reference and also in Ref.~\cite{Kanzi:2022vhp}.

Although there are plenty of examples to consider,\footnote{See, for instance, the method to find a specific type of vacuum solutions from GR described in Sect.~7.2 of Ref.~\cite{Adler}.} we believe those two, along with their own range of possibilities, are enough to reveal the simplicity and potential of the developed method to provide new solutions to this Lorentz symmetry breaking model.

\section{Conclusions and remarks}
\label{conclusions}

A great mathematical simplification was provided to the vacuum field equations of this bumblebee-gravity scenario when a gradient bumblebee field at its VEV is considered. Despite being a restriction among the possibilities for this field, some very interesting results were obtained.

After showing the pair of dynamical equations reduce to only a simple one, in this case, it was considerably simplified even further when written in terms of the background metric (\ref{background metric (xi explicit)}), a conveniently defined combination of the metric tensor and the bumblebee field. In terms of it, the dynamical equation becomes exactly the same as Einstein's equation in vacuum. As the background metric also corresponds precisely to what the metric would be in the absence of the bumblebee field, its name is properly justified thus.

A rather impressive result was obtained after writing the metric in terms of the background metric and the bumblebee field (\ref{metric in terms of the background metric}), as the resulting expression splits the spacetime description into the former plus a contribution from the latter, which merely consists of a single additive term. This is actually remarkable, as it explicitly reveals how exactly spacetime is modified by the presence of that field.

With respect to the specific conditions for it, in turn, only the fact of being at its VEV does depend on the spacetime metric, as it consists of having a certain constant modulus (\ref{const bumblebee field modulus}). However, this equivalently implies a constant modulus with regard to the background metric as well (\ref{bumblebee modulus via background metric}). Being a gradient, on the other hand, is an intrinsic characteristic, i.e., it has no relation with the spacetime metric whatsoever (\ref{null bumblebee field strength}).

As a consequence, the task of finding solutions to this model is immensely simplified. Any vacuum metric from GR according to which a gradient bumblebee field possesses a constant modulus suffices to provide a solution. It is just a matter of taking this metric as the background metric and setting the appropriate value for the field modulus. Then, the desired solution is readily obtained from Eq.~(\ref{metric in terms of the background metric}), describing the resulting effect of the bumblebee field on spacetime.

We point out that, since the bumblebee field conditions of being a gradient with constant modulus are somewhat restrictive, one should have no expectations concerning any specific spacetime aspect of the results. The best way to deal with this particular scenario is actually to take a vacuum solution to Einstein's equation and, then, find out how that field changes it regardless of the consequences. As a matter of fact, from a pragmatic point of view, it is precisely the modifications the bumblebee field brings to a certain spacetime one is actually seeking and, in this sense, we surely provided the most simplifying approach to that. The only point is that, in this particular case, one cannot generally ascribe any specific spacetime attribute to the solution {\em a priori}.

As some examples, the developed method was implemented to investigate the changes induced in the Schwarz\-schild solution and a rotating vacuum spacetime.

In the first situation, a previously published solution \cite{Casana:2017jkc} was readily recovered and a slightly more general possibility was also very easily obtained afterwards. It was shown both provide essentially the same spacetime, as they only differ on the expression for a specific constant. The supposedly most general possibility, which gives rise to cross terms on the spacetime interval and contains all the previous solutions as particular cases, was suggested next.

For the example of a rotating background, the most general bumblebee field compatible with this approach also introduces new off-diagonal terms on the spacetime metric and all solutions from the previous example are recovered as particular cases. If the rotating feature is to be kept, however, there is no possibility at all to avoid at least one new cross term to the spacetime interval. A particular choice for the arbitrary constants was considered and it provided a very simple and interesting solution. It easily recovers Kerr spacetime just by removing the bumblebee field contributions and readily recovers the first and already published solution \cite{Casana:2017jkc} as the particular case in which there is no rotation.

Although all these examples, except for the first one that has already been published, constitute new solutions to the model, no further development was made to them. The main reason is that our intent was basically to illustrate how the method can be applied, leaving their due analysis for future publications, especially because of the complexity most of them present. Concerning the simplest ones, which are the first two solutions, the second one provides essentially the same spacetime as the first, for which many studies have already been made \cite{Kanzi:2019gtu,Ovgun:2018ran,Yang:2018zef,Li:2020wvn,DCarvalho:2021zpf,Oliveira:2021abg,Kumar_Jha_2021,Gomes:2018oyd,Ovgun:2019ygw}. Only the last example had a small further analysis with regard to its distinguished surfaces. The conclusion is that they are still given by the same expressions of their Kerr solution analogous. Consequently, the first two examples also share with Schwarzschild solution the same expressions for them, which are all the same in this case.

As it became clear from those examples, there are some arbitrariness in the solutions that actually preclude any possibility of determining or constraining the theory's parameters from observational data, as it happens with condition (\ref{constraint on ell prime}) for instance. In fact, even if there was no arbitrariness at all, that possibility would be compromised already. As one can see from Eq.~(\ref{metric in terms of the background metric}), the spacetime changes resulting from the bumblebee field are given only in terms of the product $\sqrt{\xi}B_\mu$, making it impossible to constrain $\xi$ separately from $B_\mu$. That would only be possible if at least one of the conditions (\ref{bumblebee field constraints}) was discarded.

Even though we considered only a particular case of the whole model (\ref{model action}), we believe we were extremely successful as to provide the answer to the main question of how a bumblebee field modifies a certain vacuum spacetime, since the solution was explicitly given in terms of both of them. Besides, as this treatment deals with vacuum spacetimes, it can be applied to a truly wide range of situations.

Substantial mathematical development was certainly\linebreak made to the model. Many of its different aspects and consequences can now be much more easily investigated, which may help not only to highlight its benefits but also to properly address its problems, if there are.

For all these reasons, we consider this approach a great contribution to the theory.

\begin{acknowledgements}
	We thank FAPEMA (Brazilian agency) for its financial support.
\end{acknowledgements}

\appendix

\section{The Ricci tensor in terms of $\mathbf{\tilde{g}_{\mu\nu}}$}
\label{app: The Ricci tensor}

Writing the field equation (\ref{Special field equations - Ricci (simplified)}) in terms of $\tilde{g}_{\mu\nu}$ is basically a straightforward computation. Nevertheless, we shall detail the main steps needed to achieve so in this appendix.

First, let us define, according to (\ref{background metric definitions}),
\begin{subequations}\label{metric differences}
	\begin{align}
		\Delta g_{\mu\nu} & \equiv g_{\mu\nu} - \tilde{g}_{\mu\nu} = \frac{\xi}{1 + \xi B^2}B_\mu B_\nu \text{,} \label{covariant metric difference} \\
		\Delta g^{\mu\nu} & \equiv g^{\mu\nu} - \tilde{g}^{\mu\nu} = -\xi B^\mu B^\nu \text{.} \label{contravariant metric difference}
	\end{align}
\end{subequations}
The affine connection, thus, may be rewritten as
\begin{align}
	\Gamma^\alpha_{\mu\nu} = & \frac{1}{2}g^{\alpha\beta}\left[ \partial_\mu \tilde{g}_{\beta\nu} + \partial_\nu \tilde{g}_{\beta\mu} - \partial_\beta \tilde{g}_{\mu\nu} \right. \nonumber \\ & \left. + \partial_\mu (\Delta g_{\beta\nu}) + \partial_\nu (\Delta g_{\beta\mu}) - \partial_\beta (\Delta g_{\mu\nu}) \right] \text{.}
	\label{background connection (preliminary 01)}
\end{align}
Now, making the definition
\begin{equation}
	\tilde{\Gamma}^\gamma_{\mu\nu} \equiv \frac{1}{2}\tilde{g}^{\gamma\delta}( \partial_\mu \tilde{g}_{\delta\nu} + \partial_\nu \tilde{g}_{\delta\mu} - \partial_\delta \tilde{g}_{\mu\nu} ) \text{,}
	\label{Christoffel symbol for the background metric}
\end{equation}
it implies
\begin{equation}
	\tilde{g}_{\gamma\beta}\tilde{\Gamma}^\gamma_{\mu\nu} = \frac{1}{2}( \partial_\mu \tilde{g}_{\beta\nu} + \partial_\nu \tilde{g}_{\beta\mu} - \partial_\beta \tilde{g}_{\mu\nu} ) \text{.}
	\label{contractec Christoffel symbol for the background metric}
\end{equation}
Moreover, we may write
\begin{align}
	\partial_\mu & \Delta g_{\beta\nu} + \partial_\nu \Delta g_{\beta\mu} - \partial_\beta \Delta g_{\mu\nu} \nonumber \\ & = \nabla_\mu \Delta g_{\beta\nu} + \nabla_\nu \Delta g_{\beta\mu} - \nabla_\beta \Delta g_{\mu\nu} + 2\Gamma^\gamma_{\mu\nu}\Delta g_{\beta\gamma}
	\label{background connection diff derivative}
\end{align}
and, then, (\ref{background connection (preliminary 01)}) turns into
\begin{align}
	\Gamma^\alpha_{\mu\nu} = & g^{\alpha\beta}\tilde{g}_{\gamma\beta}\tilde{\Gamma}^\gamma_{\mu\nu} + g^{\alpha\beta}\Gamma^\gamma_{\mu\nu}\Delta g_{\beta\gamma} \nonumber \\ & + \frac{1}{2}g^{\alpha\beta}( \nabla_\mu \Delta g_{\beta\nu} + \nabla_\nu \Delta g_{\beta\mu} - \nabla_\beta \Delta g_{\mu\nu} ) \text{.}
	\label{background connection (preliminary 02)}
\end{align}

Passing the $\Gamma^\gamma_{\mu\nu}$ term on the right-hand side to the left and contracting the equation with $g_{\alpha\rho}$ gives
\begin{align}
	g_{\alpha\rho}\Gamma^\alpha_{\mu\nu} - \Gamma^\gamma_{\mu\nu}\Delta g_{\rho\gamma} = & \tilde{g}_{\gamma\rho}\tilde{\Gamma}^\gamma_{\mu\nu} \nonumber \\ & + \frac{1}{2}( \nabla_\mu \Delta g_{\rho\nu} + \nabla_\nu \Delta g_{\rho\mu} - \nabla_\rho \Delta g_{\mu\nu} ) \text{.}
	\label{background connection (preliminary 03)}
\end{align}
The left-hand side is just $\tilde{g}_{\rho\gamma}\Gamma^\gamma_{\mu\nu}$, thus, contraction with $\tilde{g}^{\alpha\rho}$, finally, provides
\begin{subequations}\label{connection difference}
	\begin{gather}
		\Gamma^\alpha_{\mu\nu} = \tilde{\Gamma}^\alpha_{\mu\nu} + \Delta\Gamma^\alpha_{\mu\nu} \label{background connection} \text{,} \\
		\Delta\Gamma^\alpha_{\mu\nu} \equiv \frac{1}{2}\tilde{g}^{\alpha\beta}( \nabla_\mu \Delta g_{\beta\nu} + \nabla_\nu \Delta g_{\beta\mu} - \nabla_\beta \Delta g_{\mu\nu} ) \text{.}
		\label{connection difference (metric difference)}
	\end{gather}
\end{subequations}
From definition (\ref{covariant metric difference}) and with the help of equations (\ref{Ricci simplifications}) and (\ref{Differential consequences for special B}), it is straightforward to obtain
\begin{equation}
	\Delta\Gamma^\alpha_{\mu\nu} = \xi B^\alpha \nabla_\mu B_\nu \qquad \therefore \qquad \Delta\Gamma^\alpha_{\mu\alpha} = 0 \text{.} \label{specific connection difference}
\end{equation}

The Ricci tensor, in turn, can be written in terms of its analogous $\tilde{R}_{\mu\nu}$ as
\begin{subequations}\label{Ricci difference}
	\begin{align}
		R_{\mu\nu} = & \; \tilde{R}_{\mu\nu} + \Delta R_{\mu\nu} \label{background Ricci} \text{,} \\
		\Delta R_{\mu\nu} \equiv & \; \nabla_\alpha \Delta\Gamma^\alpha_{\mu\nu} - \nabla_\nu\Delta\Gamma^\alpha_{\mu\alpha} \nonumber \\ & + \Delta\Gamma^\beta_{\mu\alpha} \Delta\Gamma^\alpha_{\beta\nu} - \Delta\Gamma^\beta_{\mu\nu} \Delta\Gamma^\alpha_{\beta\alpha} \text{.} \label{Ricci difference (connection difference)}
	\end{align}
\end{subequations}
Only the first term on the right-hand side of the last expression remains since all the others vanish as a consequence of (\ref{Differential consequences for special B}) and (\ref{specific connection difference}). Thus, the Ricci tensor finally reads
\begin{equation}
	R_{\mu\nu} = \tilde{R}_{\mu\nu} + \xi \nabla_\alpha ( B^\alpha \nabla_\mu B_\nu ) \text{.}
\end{equation}

\section{Relation between $\mathbf{g}$ and $\mathbf{\tilde{g}}$}
\label{app: metric determinant}

In order to obtain the relation between $g$ and $\tilde{g}$, recall that, for any $A_{\alpha\beta}$, its determinant, $A$, is given by either of these expressions: (see p. 18 of Ref.~\cite{landau1975classical})
\begin{subequations}\label{determinant definitions}
	\begin{align}
		A & = \frac{1}{24}\epsilon^{\alpha\beta\gamma\delta}\epsilon^{\kappa\lambda\mu\nu} A_{\alpha\kappa}A_{\beta\lambda}A_{\gamma\mu}A_{\delta\nu} \text{,} \label{isolated determinant (generic matrix)} \\
		\epsilon_{\alpha\beta\gamma\delta} A & = -\epsilon^{\kappa\lambda\mu\nu} A_{\alpha\kappa}A_{\beta\lambda}A_{\gamma\mu}A_{\delta\nu} \text{,} \label{Levi-Civita determinant (generic matrix)}
	\end{align}
\end{subequations}
in which $\epsilon^{\alpha\beta\gamma\delta}$ is the completely anti-symmetric Levi-Civita symbol so that $\epsilon^{0123} = 1$ whereas $\epsilon_{\alpha\beta\gamma\delta}$ presents the same anti-symmetries but $\epsilon_{0123} = -1$ instead. From these definitions, it follows that
\begin{equation}
	\epsilon^{\alpha\beta\gamma\delta}\epsilon_{\kappa\beta\gamma\delta} = -6\delta^\alpha_\kappa \text{,}
	\label{Levi-Civita contraction}
\end{equation}
which can be used to pass from (\ref{Levi-Civita determinant (generic matrix)}) to (\ref{isolated determinant (generic matrix)}).

If $A \neq 0$, then (\ref{Levi-Civita determinant (generic matrix)}) and (\ref{Levi-Civita contraction}) allow one to check that
\begin{equation}
	A^{\delta\nu} \equiv \frac{1}{6A}\epsilon^{\alpha\beta\gamma\delta}\epsilon^{\kappa\lambda\mu\nu}A_{\alpha\kappa}A_{\beta\lambda}A_{\gamma\mu}
	\label{generic matrix inverse}
\end{equation}
is the inverse of $A_{\rho\nu}$, i.e., $A^{\delta\nu}A_{\rho\nu} = \delta^\delta_\rho$.

Applying (\ref{isolated determinant (generic matrix)}) for $g$ and making use of (\ref{metric in terms of the background metric}), there results
\begin{equation}
	g = \tilde{g} + \frac{k}{6}\epsilon^{\alpha\beta\gamma\delta}\epsilon^{\kappa\lambda\mu\nu} \tilde{g}_{\alpha\kappa}\tilde{g}_{\beta\lambda}\tilde{g}_{\gamma\mu}B_\delta B_\nu \text{,}
	\label{background metric determinant with Levi-Civita}
\end{equation}
with $k \equiv \xi/(1 + \xi B^2)$. No higher power of $k$ remains since all of them have products of $B_\mu$ whose indices are contracted with the same Levi-Civita symbol, giving zero thus. Besides that, all the four terms in $k$ are equal, justifying the denominator equal to six.

From (\ref{generic matrix inverse}) applied to $\tilde{g}_{\mu\nu}$, this last result can be rewritten as
\begin{equation}
	g = \tilde{g} + k\,\tilde{g}\,\tilde{g}^{\delta\nu}B_\delta B_\nu = \tilde{g}\Bigl(1 + k \tilde{B}^2\Bigr) = \tilde{g}\Bigl(1+\xi B^2 \Bigr) \text{,}
	\label{background metric determinant demonstrated}
\end{equation}
in which (\ref{bumblebee modulus via background metric}) have been used.

\section{Distinguished surfaces of the particular rotating spacetime solution}
\label{app: IRS and horizons}

In this appendix, both the infinite redshift surfaces and the horizons associated with the spacetime solution described by (\ref{Kerr-like metric solution}) will be determined.

The infinite redshift surfaces are those at which $g_{00}$ is zero \cite{Adler} and are, thus, given by:
\begin{equation}
	r_\infty^\pm = M \pm \sqrt{M^2 - a^2\cos^2\theta} \text{.}
	\label{infinite redshift surfaces app}
\end{equation}
This is precisely the same expression provided by the Kerr solution, as it could not be otherwise, since the bumblebee field does not change that metric component.

The horizons, in turn, are any surface whose normal vector is null \cite{Adler}. As there is no time or axial dependence on spacetime, the most general possibility is written as:
\begin{equation}
	u(r,\theta) = 0
	\label{general horizon possibility} \text{.}
\end{equation}
Any normal vector to this surface is proportional to the gradient:
\begin{equation}
	n_\alpha = \partial_\alpha u = \left(0, \frac{\partial u}{\partial r},\frac{\partial u}{\partial \theta}, 0\right) \text{,}
	\label{general horizon normal vector}
\end{equation}
so, a necessary and sufficient condition for that surface to be a horizon is:
\begin{align}
	n^\alpha n_\alpha = & \left(\rho^2 + \ell r^2 \right)\left(\frac{\partial u}{\partial \theta}\right)^2 - 2a\ell r\sqrt{\Delta}\cos\theta\,\frac{\partial u}{\partial r} \frac{\partial u}{\partial \theta} \nonumber \\ & + \Delta\left(\rho^2 + a^2 \ell\cos^2\theta\right)\left(\frac{\partial u}{\partial r}\right)^2 = 0 \text{.}
	\label{null general normal vector condition}
\end{align}
This is clearly different from its Kerr solution analogous, which corresponds to setting $\ell = 0$ in this equation, but the final result is still the same though, as will be seen.

Firstly, just for simplicity, let us consider that one of those partial derivatives is null, i.e., $u = u(r)$ or $u = u(\theta)$. The first case would imply, according to Eq.~(\ref{null general normal vector condition}):
\begin{align}
	\Delta&\Bigl(\rho^2 + a^2 \ell\cos^2\theta\Bigr)\left(\frac{\partial u}{\partial r}\right)^2 \nonumber \\ & = \Delta\Bigl[r^2 + a^2(1 + \ell)\cos^2\theta\Bigr]\left(\frac{\partial u}{\partial r}\right)^2 = 0 \text{.}
	\label{spherical horizon condition}
\end{align}
The only possible solution consistent with $u = u(r)$ to this equation is:
\begin{equation}
	\Delta = 0 \quad \Leftrightarrow \quad r_\pm = M \pm \sqrt{M^2 - a^2} \text{.}
	\label{Kerr-like horizon app}
\end{equation}

The other case, in turn, would demand:
\begin{align}
	\Bigl(&\rho^2 + \ell r^2 \Bigr)\left(\frac{\partial u}{\partial \theta}\right)^2 \nonumber \\ & = \Bigl[(1 + \ell)r^2 + a^2 \cos^2\theta \Bigr]\left(\frac{\partial u}{\partial \theta}\right)^2 = 0 \text{.}
	\label{odd horizon condition}
\end{align}
However, there is no solution consistent with $u = u(\theta)$ to this equation and, therefore, this assumption should be ruled out. Actually, it would be really odd if that was possible.

Finally, it will be shown that the most general situation, in which none of $\partial_1 u$ and $\partial_2 u$ is null, i.e., $u$ actually depends on both $r$ and $\theta$, should also be discarded. To conclude this, one must first note that Eq.~(\ref{null general normal vector condition}) consists of a quadratic equation for both $\partial_1 u$ and $\partial_2 u$ whose solutions will never be real as long as $\Delta$ is different from zero.

After all, for $\partial_i u$ ($i$ equal to $1$ or $2$) to be real, the discriminant of Eq.~(\ref{null general normal vector condition}) should be nonnegative:
\begin{equation}
	-4\Delta\rho^4(1 + \ell)\left(\partial_j u\right)^2 \geqslant 0 \text{,}
	\label{condition for real horizon}
\end{equation}
with $j \neq i$ ($j$ equal to $1$ or $2$). Since, from (\ref{constraint on ell prime}), $\ell$ must be small, this implies a nonpositive $\Delta$. On the other hand, a real $\partial_i u$ also demands a nonnegative $\Delta$ because of the cross term in Eq.~(\ref{null general normal vector condition}). So, the only possibility of having a real surface is a null $\Delta$. However, this coincides with condition (\ref{Kerr-like horizon app}) associated with a surface described by $u = u(r)$ and, thus, already discards the possibility of having $u = u(r,\theta)$.

In fact, if we insisted on Eq.~(\ref{null general normal vector condition}) for a null $\Delta$, the condition for a horizon would become the same as Eq.~(\ref{odd horizon condition}), which can never be satisfied unless $\partial_2 u$ is zero, as $\ell$ is small. Hence, this last attempt should also be ruled out as already anticipated.

In conclusion, although Eq.~(\ref{null general normal vector condition}) is different from its Kerr solution analogous, both of them provide the same expressions (\ref{Kerr-like horizon app}) for their horizons.

\bibliography{references}

\end{document}